 \documentclass[modern,tighten,trackchanges]{aastex631}
\usepackage{amsmath}

\newcommand{\eb}{\begin{equation}}
\newcommand{\ee}{\end{equation}}

\usepackage{subfigure,graphicx}
\usepackage{url}
\usepackage{color}
\definecolor{darkgray}{gray}{0.4}
\definecolor{patinared}{rgb}{.72,.10,0}
\definecolor{patinablue}{rgb}{0,.20,.65} 
\definecolor{orange}{rgb}{1,0.5,0}
\definecolor{rkka}{RGB}{219,66,32}

\hypersetup{
  colorlinks=True,      
  urlcolor=darkgray,    
  linkcolor=patinared,  
  citecolor=patinablue, 
}

\shorttitle{Directions of astrometric quasar excursions}
\shortauthors{Makarov et al.}
\graphicspath{{./}{figures/}}

\begin{document}

\title{Partial Alignment of Astrometric Position Excursions of International Celestial Reference Frame Quasars with Radio Jet Structures}

\email{valeri.v.makarov.civ@us.navy.mil}

\author[0000-0003-2336-7887]{Valeri V.\ Makarov}
\affiliation{U.S. Naval Observatory, 3450 Massachusetts Ave NW, Washington, DC 20392-5420, USA}

\author[0000-0002-8736-2463]{Phil Cigan}
\affiliation{U.S. Naval Observatory, 3450 Massachusetts Ave NW, Washington, DC 20392-5420, USA}

\author[0000-0001-8009-995X]{David Gordon}
\affiliation{U.S. Naval Observatory, 3450 Massachusetts Ave NW, Washington, DC 20392-5420, USA}

\author[0000-0002-4146-1618]{Megan C.\ Johnson}
\affiliation{U.S. Naval Observatory, 3450 Massachusetts Ave NW, Washington, DC 20392-5420, USA}

\author[0000-0001-5944-9118]{Christopher DiLullo}
\affiliation{U.S. Naval Observatory, 3450 Massachusetts Ave NW, Washington, DC 20392-5420, USA}
\affiliation{Computational Physics, Inc., 8001 Braddock Rd. Suite 210, Springfield, VA 22151, USA}

\author[0000-0001-6759-5502]{S{\'e}bastien Lambert}
\affiliation{SYRTE, Observatoire de Paris, Universit\'{e} PSL, CNRS, Sorbonne Universit\'{e}, LNE, 61 avenue de l’Observatoire 75014 Paris,
France}

\begin{abstract}

Published analyses of very long baseline interferometry (VLBI) data for the sources included in the third International Celestial Reference Frame (ICRF3) catalog have revealed object-specific, excess astrometric variability and quasi-coherent trajectories as functions of time. A fraction of these sources show markedly elongated distributions of positions on the sky measured with diurnal observations. Here we apply a novel statistical and data-processing method to the diurnal position measurements stretching over 40 years to quantify the degree of elongation and its position angle, for each source with more than 200 data points. We find that 49\% of the examined sources have distribution elongations in excess of 1.3. Robust uncertainties of the directions of maximal astrometric dispersion are computed by the bootstrapping method, and the results are compared with a larger catalog of radio jet directions by \citet{2022ApJS..260....4P}. Nearly one-half of the sources with smaller position angle uncertainties are found to have astrometric position excursions from their mean positions aligned with the radio jet structures within $\pm 30\degr$.

\end{abstract}


\section{Introduction} \label{int.sec}

Astrometric accuracy of Very Long Baseline Interferometry (VLBI) measurements and intrinsic positional stability of radio-loud quasars are of paramount importance for the temporal consistency and rigidity of the International Celestial Reference Frame \citep[ICRF;][]{1998AJ....116..516M}, which is currently adopted by the International Astronomical Union as the realization of the International Celestial Reference System \citep[ICRS; e.g.,][]{1995A&A...303..604A} at radio wavelengths. The latest version, ICRF3 \citep{2020A&A...644A.159C}, adopted as the fundamental realization of the ICRS in 2019,\footnote{\url{https://www.iau.org/static/resolutions/IAU2018_ResolB2_English.pdf}} is defined not only in the previously utilized $S$/$X$ bands (2.3/8.4~GHz, respectively), but also in the $K$ and $X$/$Ka$ bands (24~GHz and 8.4/32~GHz, respectively). The $S$/$X$ catalog remains the largest and the most observed component, including over 5000 radio sources \footnote{in the most recent USNO global solutions\url{https://crf.usno.navy.mil/quarterly-vlbi-solution}}. These distant extragalactic objects (radio-loud AGNs) have been assumed to be extremely compact and astrometrically stable sources of light. Evidence has been recently presented, however, that this is not true for some of these sources \citep{2020RNAAS...4..108T, 2023AJ....165...69T}. In a broader statistical study of more than 5500 radio sources (including 4536 ICRF3 objects), \citet{2024arXiv240801373C} found that practically all quasars with more than 200 single-epoch precision measurements show excessive dispersion around their long-term mean positions on the sky, which is not adequately represented by the available formal covariance. The position residuals do not follow the expected bivariate normal distribution. The sample distribution has powerful tails stretching well beyond the formal confidence limits. Furthermore, when the sequence of single-epoch positions is viewed as a time series, some of the well-observed sources show quasi-coherent trajectories upon suitable smoothing of the residuals. 

The apparent astrometric instability of ICRF radio sources is probably related to the misalignment of individual positions as a function of wavelength, which becomes apparent when the VLBI-based astrometry is compared with the optical mean positions measured by the ESA Gaia mission \citep{2016A&A...595A...1G}. At the level of precision achieved so far on both sides, at least one-third of common radio-optical sources show statistically significant offsets outside the acceptable confidence levels \citep{2017A&A...598L...1K,2017ApJ...835L..30M,2017MNRAS.467L..71P,2018AJ....155..229F,2019ApJ...873..132M,2019MNRAS.482.3023P,2020A&A...634A..28L}. Multiple, object-specific factors can contribute to this positional discrepancy on both the optical and radio sides. A significant fraction of the large radio-optical offsets was found to correlate with the directions of detected linear, jet-like radio features \citep{2017A&A...598L...1K,2019MNRAS.482.3023P,2019ApJ...871..143P}, as well as with more generally evaluated radio source structures \citep{2021A&A...647A.189X}. The synchrotron opacity of the compact radio core can change on timescales from days to decades resulting in a so-called ``core shift'' variability \citep[e.g.,][]{2019MNRAS.485.1822P} at the $0.1-1$~mas level \citep{2011A&A...532A..38S, 2012A&A...545A.113P}.

In this paper, we look at the problem of frequency-dependent and time-variable astrometric positions of ICRF3 sources from a different angle, employing the available single-epoch precision VLBI data to determine the position angles of the intrinsic wobbles on the sky. The general idea is related to the previous work by \citet{2021A&A...648A.125G}, who demonstrated that the vectors of position offsets for each source are often (in more than half of the sample) not uniformly distributed in position angle. Some of the ICRF objects were found to present a single ``preferred" direction of astrometric dispersion. However, the majority of these cases were found to be consistent with enhanced dispersion in the local south-north direction, i.e., in the declination coordinate. The relative under-performance of VLBI astrometry in declination is a fully expected consequence of the geometric configuration of the typical VLBI session networks and the intrinsic atmosphere calibration uncertainties. The baselines involved in each session are mostly east-west oriented, creating more favorable conditions to measure the R.A. coordinate component with higher precision. This geometric asymmetry should be captured by the formal covariance of each single-epoch position measurement, reflecting in the error ellipses elongated mostly in the south-north direction on the sky. Our goal is to find a method that is impervious to the expected stochastic asymmetry of single-epoch VLBI astrometry, allowing us to reveal the physical orientation of apparent position excursions.

\section{Data} \label{data.sec}
We use a data set of diurnal VLBI sessions spanning over 40 years of continued observations with the global IVS and VLBA networks. The data are taken from the usn2023a USNO global solution, which is described in more detail by \citet{2024arXiv240801373C}. The principles of the VLBI global solutions can be found in \citep{2022Univ....8..374D}. The standard simultaneous S/X band setup is utilized, where the S-band (2.3~GHz) is used for determining the ionosphere contribution to the delay, and improving the final solution defined at 8.4~GHz in the X-band. The usn2023a solution provides a time sequence of individual (RA, Decl.) coordinates for each single session of VLBI measurements, which nominally correspond to one day. Each source has a number of individual position determinations ranging from a few to a few thousands for the most frequently observed objects. The mean positions over the entire span of VLBI observations can be computed and compared with the mean positions tabulated in the ICRF3 S/X catalog.

The input data for this analysis are the time series from 6581 of these diurnal sessions. Each data point is a tuple of R.A. and decl. coordinates, which are defined in the ICRF3. The coordinates have three associated formal covariance parameters, including the standard errors (uncertainties) in mas and the correlation coefficient. Closely following the procedure described in \citet{2024arXiv240801373C}, we start by computing the mean position for each source using the Maximum Likelihood (ML) principle and assume that the coordinate measurements are normally distributed with the specified covariances. The residual offset vectors are then computed for each session and each source by subtracting the mean positions computed in the ML framework. For practical convenience, these offset vectors are called $[x,y]^T$ here, denoting the local residuals $\Delta\alpha\cos{\delta}$ and $\Delta\delta$, respectively. The corresponding formal covariance matrix is designated as $\boldsymbol{C}_{xy}$.

Our analysis is further restricted to 265 ICRF3 sources with at least 200 individual observing sessions each. This filter is driven by the required number statistics to obtain a reliable determination of possible asymmetry in the scatter of individual offsets. Approximately half of the selected objects are defining ICRF3 sources, which represent the backbone of this celestial frame, because they are internally used to align and calibrate the global VLBI solutions. The number of measurements is uneven among the 265 sources, with some of them counting up to a few thousands data points.

\section{Method}
\label{method.sec}

\begin{figure*}
\includegraphics[width=0.48 \textwidth]{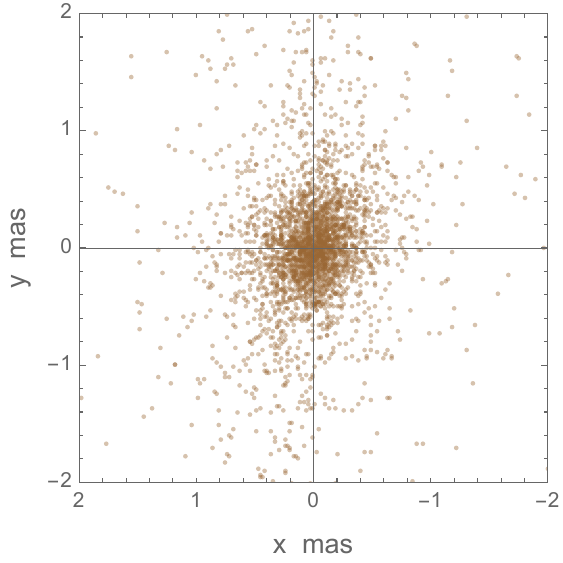}
\includegraphics[width=0.48 \textwidth]{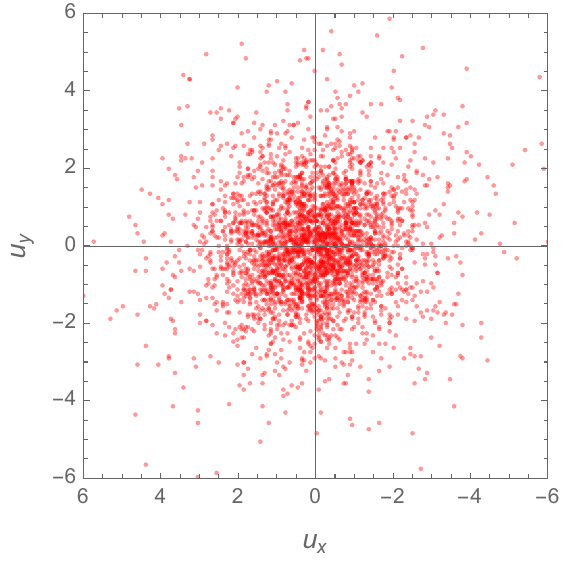}
\caption{VLBI-measured position offsets of the ICRF3 source IERS B1334$-$127 with respect to its maximum-likelihood mean position. Left: Offsets in tangential sky coordinates $x$ (Right Ascension) and $y$ (Declination) in mas. 
Right: Standardized offsets $\{u_x,u_y\}$ (Eq. \ref{u.eq}) relative to the same mean position.}
\label{xy.fig}
\end{figure*}

Our method uses the standardized offset vector
\eb
\boldsymbol{u}\equiv [u_x,u_y]^T = \boldsymbol{C}_{xy}^{-\frac{1}{2}}\;[x,y]^T,
\label{u.eq}
\ee
which is expected to follow a bivariate normal distribution of unit variance and zero correlation. In reality, these vectors invariably show excess dispersion for sources with more than 200  measurements, which we have interpreted as evidence for genuine stochastic trajectories of ICRF quasars (or ``cosmic error"). However, this operation helps to remove most of the technical distortion in the sample distribution of residuals related to the predictable covariances of coordinate measurements. An example of such improvement is shown in Fig.~\ref{xy.fig}. The left panel is the scatter plot of the VLBI measured position offsets $\{x,y\}$ for the ICRF3 quasar IERS B1334$-$127, which is one of the most-observed sources with 3174 single-epoch determinations in our database. The cloud of points is conspicuously elongated in the S-N direction, possibly with a small tilt. This shape is consistent with the formal covariances, which indicate larger uncertainties in the declination component due to the configurations of VLBI baselines. The right plot displays the distribution of standardized $\boldsymbol{u}$ endpoints according to Eq. \ref{u.eq}. This distribution is much more rotation-symmetric without any obvious direction of enhanced dispersion. Thus, in this case, the standardization effectively removes the technical deformation of position excursions, which is not related to the physics of the object.

In \citet{2024arXiv240801373C}, the normalized single-epoch position offset
\eb
D=\sqrt{[x,y]\,
\boldsymbol C_{xy}^{-1}\,[x,y]^T}=\sqrt{\boldsymbol{u}^T\,\boldsymbol{u}}
\label{D.eq}
\ee
and the related dimensionless score parameter $Q68$, which is the 68th percentile of the $D$ sample, were used to quantify the magnitude of the excessive dispersion. Here, we compute a different quantity, which is the position angle $\vartheta$ of the long axis of the possible remaining elongation of the standardized residuals $\boldsymbol{u}$, from the local North ($y$) direction.

The observed distributions of standardized offsets are often non-Gaussian and show a variety of shapes. One possible approach would be to fit a common functional form with free parameters for each source. The bivariate Student T distribution could be a reasonable choice, because it runs the gamut of heavy-tailed functions from Normal to Cauchy types depending on the fitted number of degrees of freedom. In this study, we chose an alternative non-parametric approach, which avoids possible systematic errors due to a mismatch of the model distribution. We compute the empirical distribution function of $\boldsymbol{u}$ for each source using Wolfram Mathematica's {\it EmpiricalDistribution} function\footnote{\url{https://reference.wolfram.com/language/ref/EmpiricalDistribution.html}}. The cumulative distribution function (CDF) is represented by a step function counting the number of instances below the argument point. This is a sufficiently accurate (but discretized) representation of the CDF for a large sample. The emerging CDF can be used to compute the PDF, moments (including the mean), and the 2 by 2 covariance matrix. The latter contains the information about the confidence areas (ellipses) that we need for this analysis.

If $\boldsymbol{C}_u$ is the empirical covariance matrix of the sample distribution of $\boldsymbol{u}$ for a given source, its eigenvectors $\boldsymbol{e}_1$ and $\boldsymbol{e}_2$ and the respective eigenvalues $\epsilon_1$ and $\epsilon_2$ can be directly computed. An elongated distribution along a certain direction is evident as $\epsilon_1/\epsilon_2 > 1$, in which case the position angle of the vector $\boldsymbol{e}_1\equiv [e_{1x},e_{1y}]^T$ defines this preferred direction. Thus, we compute the degree of elongation
\eb 
\varepsilon=\epsilon_1/\epsilon_2,
\label{el.eq}
\ee 
and the position angle of the major axis of the confidence ellipse by\footnote{Note that in the adopted Wolfram's notation, the two-argument $\arctan(x,y)$ corresponds to $\arctan(y/x)$, i.e., the arguments are swapped.}
\eb 
\vartheta = \arctan(e_{1y},e_{1x}).
\label{pa.eq}
\ee 
This furnishes the angle in accordance with the astronomical convention, where it is rendered from the local North direction through East. However, due to the axial symmetry of the confidence ellipse, this direction is modulo-$\pi$ invariant. The emerging negative values of $\vartheta$ are redefined to the support interval $[0\degr,180\degr]$ by adding $180\degr$.

While this calculation is relatively straightforward, it is also important to estimate the associated uncertainties. Here, in the spirit of robust statistical analysis, we use the bootstrapping method. For each source, we perform 3000 trial computations of $\boldsymbol{e}_1$, $\boldsymbol{e}_2$, $\varepsilon$,  and $\vartheta$ for randomly selected (without repetition) subsets of data points including half of the entire sample. Robust estimates of the uncertainties of $\varepsilon$,  and $\vartheta$ are computed as the difference of $0.84$- and $0.16$-quantiles of the bootstrapped sample distributions divided by $2\sqrt 2$. The latter coefficient factors in the expected loss of precision due to the halving the available data set. The resulting robust uncertainties are denominated $\sigma_\varepsilon$  and $\sigma_\vartheta$.

\section{Results}

The computed position angles of the greatest dispersion in $\boldsymbol{u}$ for 265 ICRF quasars are fairly uniform in the interval $[0\degr,180\degr]$ with a moderate enhancement at $P\simeq 90\degr$. This enhancement is likely to be the remaining artifact of the VLBI astrometry technique. The dominant orientation of the VLBI baselines in the east-west direction on the surface of Earth results in a considerably better formal precision of the measured R.A. components. The cosmic error is assumed to be isotropic, but it shows more clearly in R.A. because of the smaller formal uncertainties. We are, however, mostly interested in a subset of the 265 sources, where the angle uncertainties are small enough and the elongation $\varepsilon$ is large enough for a meaningful comparison with imaging data.

We find that the majority of the initial sample have moderate elongation estimates $\varepsilon$, with 49\% of the sample $\epsilon$ exceeding 1.3, and 29\% above 1.5. Reliable estimates can only be obtained for objects with significant elongations of $\boldsymbol{u}$-distributions. The standard errors $\sigma_\vartheta$, which quantify the robustly estimated internal precision of $\vartheta$, are smaller than $5\degr$ for 25\% of the sample, and smaller than $10\degr$ for 53\% of the sample. We note that a large $\sigma_\vartheta$ value does not necessarily signify a stable, astrometrically fixed source, but rather, absence of a specific direction of position walks.  Table~\ref{res.tab} shows the leading 10 entries of the output table with the derived parameters. The entire data sample for 265 ICRF sources is available in the online version of this manuscript.

\begin{figure}
\includegraphics[width=1\textwidth]{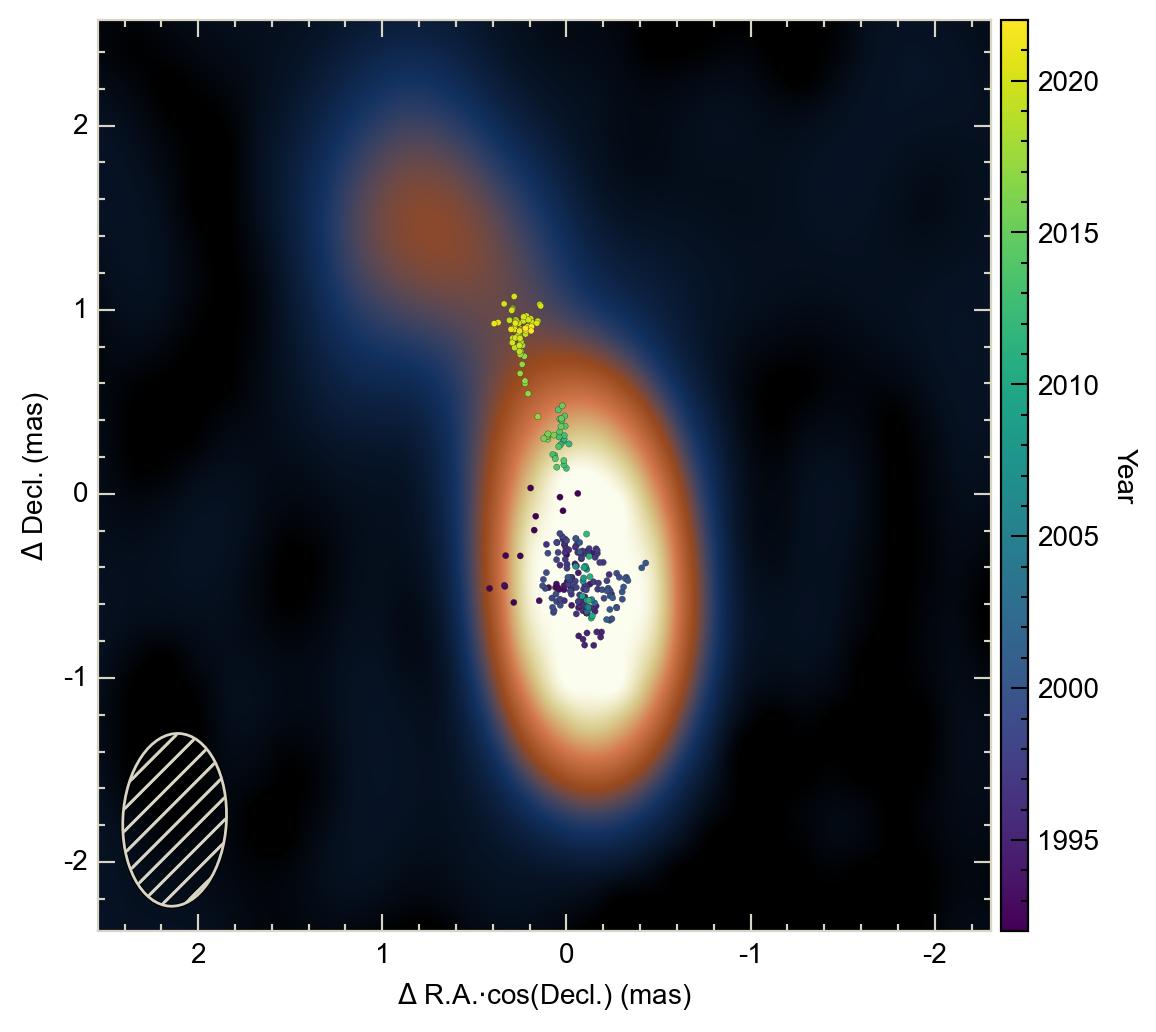}
\caption{$X$-band image of B1038+52A from USNO's FRIDA database\footnote{\url{https://crf.usno.navy.mil/FRIDA}}, observed on 2008 Jan. 23. The hatched ellipse denotes the beam FWHM. The astrometric measurements, smoothed over a rolling 4-month time window as described in \cite{2024arXiv240801373C} to enhance the visibility of coherent trends, are overlaid as scatter points colored by time. }
\label{1038+52A_X.fig}
\end{figure}

The greatest elongation $\epsilon=5.43$ is found for IVS B1038+52A = ICRF J104146.7+523328 = SDSS J104146.77+523328.2. The cloud plots of $\{x,y\}$ and $\boldsymbol{u}$ offsets similar to Fig.~\ref{xy.fig} (not displayed here for brevity) both show extremely elongated configuration, even suggesting a dual source with a $\sim 1.5$ mas separation. The computed orientation $\vartheta=16.2\pm 1.4\degr$ determined in this study is consistent with the position angle of the jet structure clearly visible in high-resolution hybrid VLBI-maps \citep{1997A&A...325..383R}. The reported {\it ibid.} changes in the separation between the sources A and B, which are unrelated quasars $33\arcsec$ apart, are undoubtedly caused by the walks of the A source, because of the aligned A-B position angle. The time evolution of the single-epoch positions clearly shows a coherent transition northward along $\vartheta$ over approximately 12 yr, illustrated in Figure~\ref{1038+52A_X.fig}.

\section{Comparison with radio jet directions}

We use the results by \citet{2022ApJS..260....4P} to compare the directions of astrometrically detected walks with position angles of radio jet structures. The input catalog includes 9220 entries, of which 228 match the objects in our working sample. The formal precision of radio-jet position angles $\vartheta_{\rm jet}$ is mostly higher (peaking at $\sim 3\degr$) than our robustly estimated precision of $\vartheta_{\rm ast}$. Therefore, we further limit this comparison to 127 ICRF sources with $\sigma_\vartheta<10\degr$ from our analysis. The jet directions are estimated on the full support interval $[-180\degr,180\degr]$, while our elongation directions are modulo-$\pi$ indeterminate. We mirrored the jet position angles by adding $180\degr$ to all negative values.

\begin{figure}
\includegraphics[width=\linewidth]{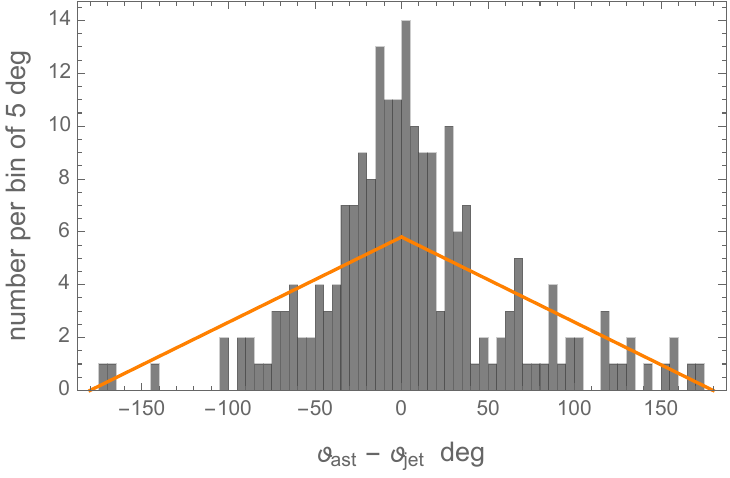}
\caption{Histogram of differences between the position angles of preferred astrometric excursions determined in this paper ($\vartheta_{\rm ast}$) and the position angles of radio jets ($\vartheta_{\rm jet}$) from \citep{2022ApJS..260....4P}, modulo $\pi$.}
\label{pp.fig}
\end{figure}

The histogram of computed $\vartheta_{\rm ast}-\vartheta_{\rm jet}$ is reproduced in Fig.~\ref{pp.fig}. The expected distribution of differences in position angles for isotropic, statistically independent samples is triangular, as shown with the orange line normalized to the total number of points and the $5\degr$ bin. The strongly bell-shaped histogram peaked around 0 indicates that a considerable fraction of VLBI sources are well-aligned with the independently estimated jet directions. Note that a few sources with P.A. differences close to $\pm 180\degr$ are also the aligned cases, because those were separated in the plot by the modulo-$\pi$ adjustment.

Counting all cases in excess of the expected distribution within the interval $[-30\degr,+30\degr]$, we estimate that nearly half of this selection have aligned directions. A slightly higher rate of alignments is found when the sample is further limited to $\epsilon>1.1$ cases. On the other end of the range, a number of sources are obviously present with significant deviations of the preferred astrometric walks from the radio-jet direction. We find 10 sources, for example, with $\vartheta_{\rm ast}-\vartheta_{\rm jet}$ in excess of $45\degr$, which have elongations $\epsilon>1.5$ and formal errors $\sigma_\vartheta<10\degr$, although two of them are cases of the $180\degr$-split. The most confidently detected case of misalignment is the blazar B2229+695, which shows a distinct quasi-coherent astrometric trajectory in R.A. The jet direction from \citep{2022ApJS..260....4P} is $43\pm 5\degr$, while our robust determination is $89.0\pm 0.9\degr$. We note, however, that the estimated core-jet shift is 2.6 mas, which is much greater than the typical astrometric excursion in R.A. within 1 mas. A small fraction of VLBI measurements show extreme outliers separated by $\sim 2$ mas and more from the mean position, whose location is more consistent with PA$\simeq 45\degr$.

\begin{figure}
\includegraphics[width=1\textwidth]{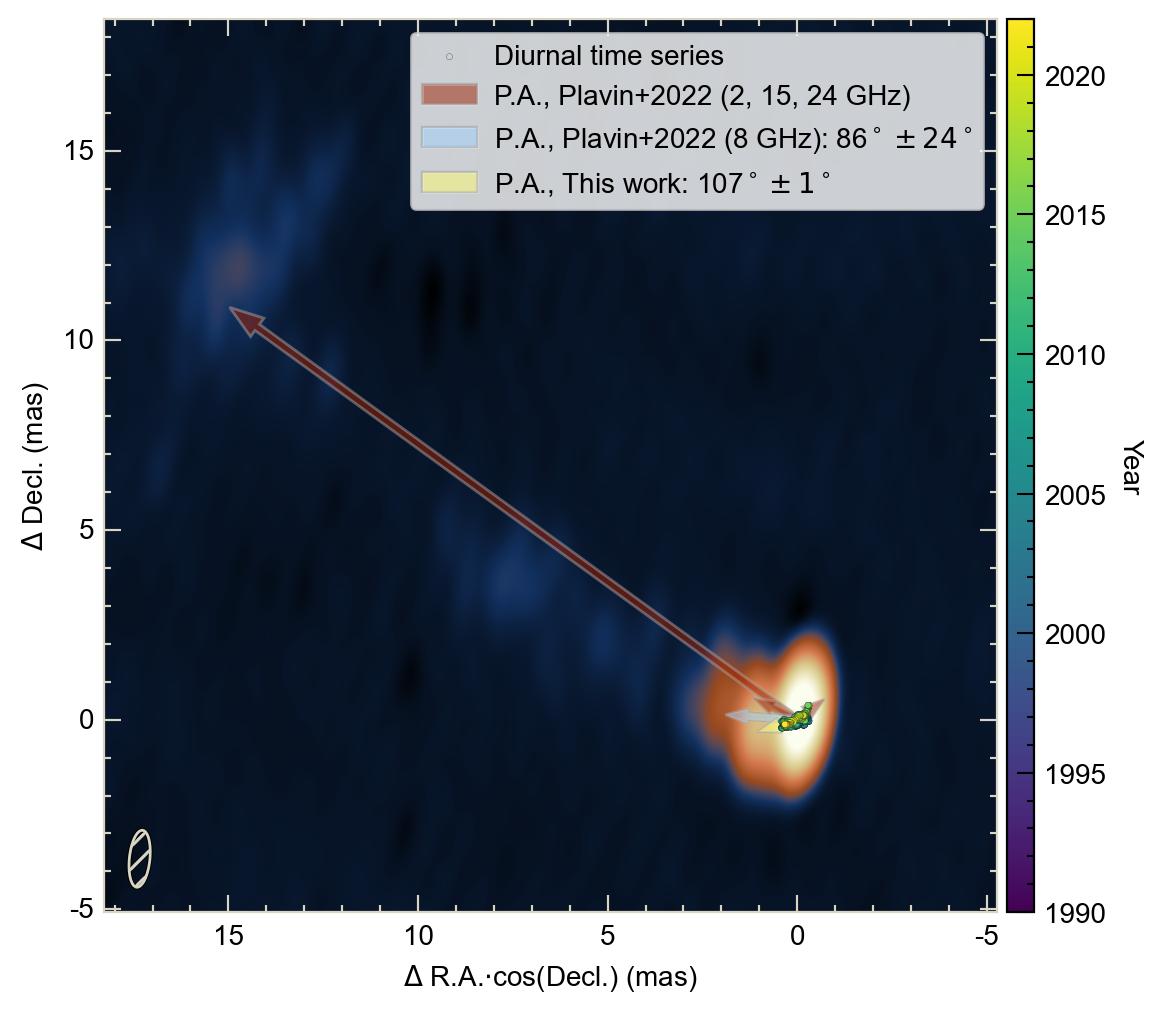}
\caption{15~GHz MOJAVE image of 0607-157 \citep{2018ApJS..234...12L}, clearly showing emission in multiple extended regions. The scatter points are the astrometric time series, colored by time, and are smoothed over a rolling 4-month time window as described in \cite{2024arXiv240801373C}.  The red arrows correspond to the jet angles and median distances determined by \cite{2022ApJS..260....4P} at various frequencies, and the blue arrow corresponds to the same at 8~GHz. The yellow arrow denotes the angle of preferred astrometric excursion $\vartheta_{\rm ast}$. }
\label{0607-157_MOJAVE.fig}
\end{figure}

The low-spectral peaked LSP quasar 0607-157 (PKS 0607-15) exhibits a northeastward extension of the jet \citep[MOJAVE 15~GHz observations suggest a stationary knot around 17~mas from the radio core,][illustrated here in Figure~\ref{0607-157_MOJAVE.fig}]{2018ApJS..234...12L}, while the inner structure within 2~mas from the core indicates rather eastward ejections. \citet{2022ApJS..260....4P} found jet directions averaging $38\pm5\degr$ across several bands, with the lowest frequencies consistent with the 17~mas extension. Our determination is $107\pm1\degr$. The difference between the two estimates suggests that our determination is more sensitive to the regions close to the VLBI centroid -- that is, in the case of 0607-157, the radio core (base of the jet) -- while \citep{2022ApJS..260....4P} estimates may be more influenced by the extended regions. Though the difference can be small for straight jets, it becomes strongly separation-dependent for curved jets. We also note the wide range of wavelength-specific determinations of jet directions and core-jet distances in \citet{2022ApJS..260....4P} for this object ranging from $50\degr$ (17.2 mas) at 1.4 GHz to $-64\degr$ (0.3 mas) at 43 GHz. Their $86\degr$ (1.3 mas) determination at 8 GHz is rather close to our astrometric data in the $X$-band in both the PA and linear extent. 

We find that the absolute difference in position angle $\vartheta_{\rm ast}-\vartheta_{\rm jet}$ appears to be significantly correlated with the viewing angle on the line of sight, the latter being available from \cite{2021ApJ...923...67H} for 110 sources in our sample. Spearman's $\rho$ is $-0.24$ with $p=0.011$ for this correlation, suggesting that, on average, small viewing angles are more often associated with greater $|\vartheta_{\rm ast}-\vartheta_{\rm jet}|$ differences. For instance, 0607-157 has a viewing angle of 0.06 deg. Jets with small viewing angles (i.e., well-aligned with the line of sight) generate statistically smaller astrometric excursions with a more isotropic distribution around the mean position. Both effects randomize the estimated position angles $\vartheta_{\rm ast}$. This interpretation finds support in the negative correlation between $|\vartheta_{\rm ast}-\vartheta_{\rm jet}|$ and $\varepsilon$ of $-0.26$ ($p<0.001$), meaning that a small elongation is associated with worse alignment in PA.

\section{Discussion and Future Work} \label{disc.sec}
The previous analysis of ICRF sources' position differences at three different radio wavelengths and in the optical \citep{2021A&A...651A..64L} supported the simple ``core-jet" model, where the respective photocenters are well aligned with the dominant radio-jet structure, but have systematic shifts along the jet. This study furnishes additional evidence that the intrinsic positional variability of ICRF sources within the single X band is often (but not always) correlated with the directions of the jets. Assuming that the projected jets are isotropic on the sky, the cosmic error component has a non-uniform impact on the stability and accuracy of the reference frame depending on specific applications. For example, the closely related Earth orientation determination and UT1$-$UTC monitoring are more sensitive to the R.A. component of reference sources' walks. Specific ICRF3 sources with significant elongations $\varepsilon$ and position angles $\vartheta$ closer to $90\degr$ should be downweighted in such solutions. The impact of anisotropic excursions on geodetic nutation series can also be predicted by means of more sophisticated modeling \citep{2008A&A...481..535L}.

Our results confirm that the alignment of astrometric walks with jet structures is not universal. A significant fraction of investigated objects show a mismatch by more than $30\degr$, which is significant with respect to the estimated uncertainty. \citet{2011AJ....141..178M} have detected this partial misalignment comparing the jet structures with ad hoc ``proper motions" of a relatively small sample. Noting that the notion of quasar proper motions based on coordinate time series is somewhat dubious, given the stochastic character of the process, we do find more evidence here that some of the ICRF quasars are not compliant with the ``core-jet" paradigm. For the example of B2229+695, we have seen a case of possible hierarchical stricture, where two different processes may be responsible for the variation of apparent position, on two angular scales $<1$ mas and $\ge 2$ mas. A possible explanation could also be found in the inverse correlation of the radio-optical position offsets with the degree of optical flux variability \citep{2022ApJ...939L..32S, 2024A&A...684A.202L}, which is interpreted as the consequence of the line-of-sight alignment of relativistic jets in blazars and other highly variable sources. Future studies can therefore take different paths. On one side, a better characterization of astrometric radio sources with large elongation $\epsilon$ is warranted, including high-resolution imaging with VLBI in different bands, if possible. The other direction of future research would concern the non-compliant sources. Are they characterized by lower degrees of optical or radio variability? Do they have prominent one-sided jets? Can any evidence be found that these jets, if they are present, rapidly change their direction on the time scale of decades?

\begin{deluxetable*}{lDD|DDDD} 
\caption{Directions of preferred astrometric walks for 265 ICRF3 quasars}
\tablehead{ \colhead{IERS name} & \multicolumn{2}{c}{mean R.A.} & \multicolumn{2}{c}{mean Decl.} & \multicolumn{2}{c}{$\vartheta$} &
\multicolumn{2}{c}{$\sigma_\vartheta$} & \multicolumn{2}{c}{$\varepsilon$} & \multicolumn{2}{c}{$\sigma_\varepsilon$} \\
 \colhead{} & \multicolumn{2}{c}{deg} & \multicolumn{2}{c}{deg} & \multicolumn{2}{c}{deg} &
\multicolumn{2}{c}{deg} & \multicolumn{2}{c}{ } & \multicolumn{2}{c}{ } }
\decimals
\startdata
 \text{0003-066} & 1.557887101 & -6.393148984 & 170.5 & 12.9 & 1.11 & 0.044 \\
 \text{0008-264} & 2.755194559 & -26.209271393 & 117.1 & 20.2 & 1.282 & 0.122 \\
 \text{0014+813} & 4.285310567 & 81.58559331 & 177.7 & 1.2 & 3.448 & 0.249 \\
 \text{0016+731} & 4.940777101 & 73.458337975 & 97.7 & 1.4 & 3.051 & 0.166 \\
 \text{0017+200} & 4.907726976 & 20.362679197 & 79.7 & 32.6 & 1.109 & 0.08 \\
 \text{0019+058} & 5.635171635 & 6.134519947 & 93. & 38.6 & 1.076 & 0.075 \\
 \text{0048-097} & 12.672155741 & -9.484781904 & 80.9 & 15.2 & 1.089 & 0.043 \\
 \text{0048-427} & 12.78959093 & -42.442581219 & 146.8 & 18.2 & 1.366 & 0.211 \\
 \text{0059+581} & 15.690676528 & 58.40309348 & 119.4 & 34.3 & 1.021 & 0.026 \\
 \text{0104-408} & 16.687948563 & -40.572211249 & 135.2 & 14.8 & 1.109 & 0.052 
\enddata
\tablecomments{
IERS name in column 1 is obtained by prepending IERS B to the given code. R.A. and decl. coordinates in columns 2 and 3 are the VLBI mean coordinates. Position angles of preferred astrometric walks $\vartheta$ in column 4 are computed per Eq. \ref{pa.eq}, and the elongation parameters $\varepsilon$ in column 6 per Eq. \ref{el.eq}, 
(see Sec.~\ref{method.sec}).
}

\label{res.tab}
\end{deluxetable*}

\section{Acknowledgments}
This work supports USNO's ongoing research into the celestial reference frame and geodesy. 
The National Radio Astronomy Observatory is a facility of the National Science Foundation operated under cooperative agreement by Associated Universities, Inc. 
The authors acknowledge use of the Very Long Baseline Array under the U.S.\ Naval Observatory’s time allocation.  
This research has made use of data from the MOJAVE database that is maintained by the MOJAVE team \citep{2018ApJS..234...12L}. The authors have used the IVS data archive maintained by the International VLBI Service for Geodesy and Astronomy, \url{https://ivscc.gsfc.nasa.gov/products-data/index.html}.

\vspace{5mm}
\facilities{VLBA}


\software{
    \texttt{Mathematica} \citep{Mathematica}, 
    }






\bibliography{main}{}
\bibliographystyle{aasjournal}



\end{document}